\documentclass[letterpaper,conference]{IEEEtran}
\IEEEoverridecommandlockouts
\usepackage{amsmath, bm, amssymb, algpseudocode, algorithm, mathtools, svg,ifluatex, pgfplots, amsthm, caption, subcaption, tikz}
\usepackage{cite, bbm, accents, url}
\usepackage{graphicx}
\usepackage{textcomp}
\usepackage{xcolor}
\usepackage{stfloats,cuted}
\usepackage[acronym,shortcuts]{glossaries}
\usepackage[letterpaper, bottom=1.05in, top=0.75in, left=0.63in, right=0.63in]{geometry}
\usetikzlibrary{graphs, positioning, quotes, shapes.geometric, arrows}

\pgfplotsset{compat=1.15}

\newacronym{isac}{ISAC}{integrated sensing and communications}
\newacronym{ofdm}{OFDM}{orthogonal frequency-division multiplexing}
\newacronym{ps}{PS}{processing system}
\newacronym{pl}{PL}{programmable logic}
\newacronym{pdsch}{PDSCH}{physical data shared channel}
\newacronym{ra}{RA}{range-angle}
\newacronym{mts}{MTS}{multi-tile synchronization}

\theoremstyle{definition}

\tikzset{%
    smallblock/.style={draw, fill=white, minimum height=1em, minimum width=1em},
    block-common/.style={draw, fill=white, minimum height=1.5em, minimum width=4em},
    block/.style={rectangle, block-common, align=center},
    txtblock/.style={block, align=center, minimum height=4em},
    bigblock/.style={block, minimum height=8em},
    txtbigblock/.style={bigblock, align=center},
    input/.style={inner sep=1pt},       
    output/.style={inner sep=1pt},      
    sum/.style = {draw, fill=white, circle, minimum size=1.1em, inner sep=0pt,
      font={\small$+$}},
    prod/.style = {draw, fill=white, circle, minimum size=1.1em, inner sep=0pt,
      font={\normalsize$\times$}},
    pinstyle/.style = {pin edge={to-,thin,black}}
}

\begin{document}
\title{Real-Time Multi-Target Detection and Tracking with mmWave 5G NR Waveforms on RFSoC
\thanks{X. Li and V. C. Andrei, and U. J. M\"onich were supported in part by the German Federal Ministry for Research, Technology and Space (BMFTR) in the programme of ``Souver\"an. 
Digital. Vernetzt." within the research hub 6G-life under Grant 16KISK002.
H. Boche was supported in part by the BMFTR Quantum Programm QDCamNetz, Grant 16KISQ077, and QUIET, Grant 16KISQ093.
    }
}

\author{\IEEEauthorblockN{Xinyang Li\IEEEauthorrefmark{1},
Hian Zing Voon\IEEEauthorrefmark{1},
Vlad C. Andrei\IEEEauthorrefmark{1}, 
Alexander Sessler\IEEEauthorrefmark{2},\\
Nunzio Sciammetta\IEEEauthorrefmark{2},
Ullrich J. M{\"o}nich\IEEEauthorrefmark{1},
Dominic A. Schupke\IEEEauthorrefmark{2} and Holger Boche\IEEEauthorrefmark{1}
}
\IEEEauthorblockA{\IEEEauthorrefmark{1}Chair of Theoretical Information Technology, Technical University of Munich, Munich, Germany\\
\IEEEauthorrefmark{2}Central Research and Technology, Airbus, Ottobrunn/Munich, Germany\\
Email: \{xinyang.li, hianzing.voon, vlad.andrei, moenich, boche\}@tum.de, \\
\{alexander.sessler, nunzio.sciammetta, dominic.schupke\}@airbus.com}
}

\maketitle

\allowdisplaybreaks
\glsdisablehyper

\begin{abstract}
We demonstrate a real-time implementation of multi-target detection and tracking using 5G New Radio (NR) physical downlink shared channel (PDSCH) waveform with 400 MHz bandwidth at 28 GHz carrier frequency. The hardware platform is built on a radio frequency system-on-chip (RFSoC) 4x2 board connected with a pair of Sivers EVK02001 mmWave beamformers for transmission and reception. The entire sensing transceiver processing and fast beam control are realized purely in the \ac{pl} part of the RFSoC, enabling low-latency and fully hardware-accelerated operation. The continuously acquired sensing data constitute 3D \ac{ra} tensors, which are processed on a host PC using adaptive background subtraction, cell-averaging constant false alarm rate (CA-CFAR) detection with density-based spatial clustering of applications with noise (DBSCAN) clustering, and extended Kalman filtering (EKF), to detect and track targets in the environment. Our software-defined radio (SDR) testbed integrates heterogeneous computing resources, including CPUs, GPUs, and FPGAs, thereby providing design flexibility for a wide range of tasks.
\end{abstract}


\glsresetall

\section{Introduction}\label{sec:intro}
Implementing sensing functionality using communication signals is one practical solution to \ac{isac}\cite{liu2022integrated}. To achieve high-resolution target range and angle estimation, the use of mmWave frequencies is preferred \cite{gu2025very} due to their large spectral bandwidth and compatibility with advanced beamforming techniques. However, the real-time processing of wideband signals and rapid beamsweeping pose significant challenges for conventional CPU/GPU-based software-defined radio (SDR) systems, which are often limited by latency and throughput constraints. 
Therefore, we have realized a complete sensing transceiver on an AMD radio frequency system-on-chip (RFSoC) platform, which integrates a \ac{ps}, \ac{pl}, and hardened high-speed ADC/DAC blocks. The RFSoC is connected to a pair of Sivers EVK02001 mmWave beamformers for up- and down-conversion between baseband and 28 GHz carrier frequency. Based on the generated \ac{ra} tensors, we perform cell-averaging constant false alarm rate (CA-CFAR) detection and extended Kalman filtering (EKF) for target tracking, with data association mechanisms to handle multi-target scenarios.

\section{System Setup}

\begin{figure}[h]
    \centering
    \begin{subfigure}[t]{0.5\textwidth}
        \centering
        \includegraphics[width=0.8\linewidth]{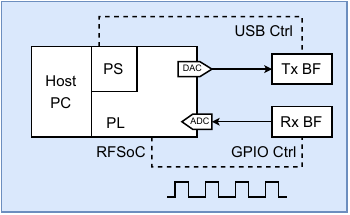}
        \caption{System architecture.}
    \end{subfigure}
    \begin{subfigure}[t]{0.5\textwidth}
        \centering
        \includegraphics[width=0.8\linewidth]{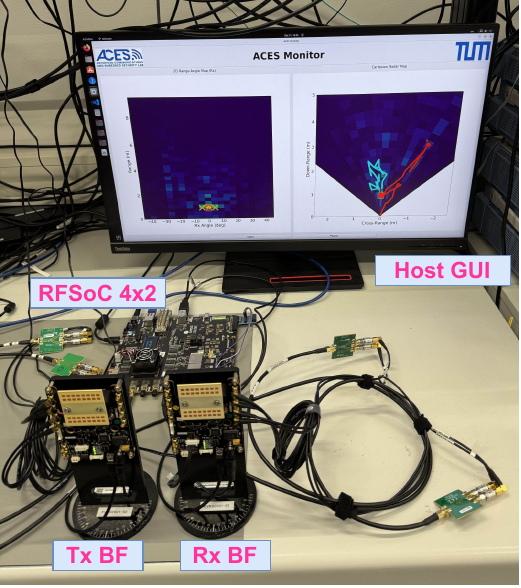}
        \caption{Real-world testbed.}
    \end{subfigure}
    \caption{Hardware setup.}
    \label{fig:hardware}
\end{figure}

The entire system comprises an RFSoC 4x2 board and two Sivers EVK02001 mmWave beamformers, one for transmission and the other for reception, as shown in Fig~\ref{fig:hardware}. The RFSoC 4x2 board serves as a baseband signal processing unit. Specifically, the AMD RFSoC architecture integrates an ARM core (\ac{ps}), an FPGA fabric (\ac{pl}), and direct RF-sampling data converters (RFDCs) into a unified system-on-chip (SoC), offering the flexibility for HW/SW codesign and high-speed RF data streaming. The Sivers EVK02001 mmWave $2\times 8$ uniform patch array (UPA) beamformer provides up- and down-conversion between baseband and Frequency Range 2 of 5G New Radio (NR FR2), supporting fast beam steering and sweeping capabilities.

We use 5G NR \ac{pdsch} as the waveform, generated by MATLAB 5G Toolbox. In particular, it occupies 275 resource blocks with subcarrier spacing 120~kHz, corresponding to a bandwidth of 400 MHz, which is the maximum allowed bandwidth at n257 frequency band. The generated baseband IQ signals are loaded from RFSoC \ac{ps} to a block RAM in \ac{pl} via AXI interface. By transmitting, the IQ samples are read sequentially from the RAM and streamed out via two DACs of the RFDC. The IQ signals are then converted to a 28 GHz carrier frequency and sent out by the Sivers beamformer.

The receiving beamformer performs the reverse operation, converting the backscattered wave into baseband IQ signals, which are then captured by RFSoC ADCs and streamed to \ac{pl}. The receiver processing steps comprise \ac{ofdm} demodulation, channel estimation, and IFFT to obtain the final range profile\cite{braun2014ofdm}. We implemented these steps on the \ac{pl} using Verilog and MATLAB HDL Coder Toolbox, enabling the real-time processing thanks to the high parallelism of FPGAs. The output range profile is saved to a block RAM on \ac{pl}, which can be read out from PS via AXI protocol. 

For each beam direction, the above steps are performed once. Therefore, by stacking the range profiles for all transmitting and receiving beams, one can obtain a 3D \ac{ra} tensor with axes range, transmit angle, and receive angle. A target in the environment may result in a spot in the \ac{ra} tensor, and the associated range and angle can be extracted. To enable a fast beam sweeping (over more than 400 beams), we control the receiving beamformer through a sequence of GPIO trigger signals generated by the RFSoC PL, where each rising edge switches the beam to the next angle. Consequently, it only takes less than 200 ms to sweep over all transmitting and receiving beams, such that the targets can be treated as quasi-static within a sweeping period.
Additionally, to synchronize the transmitter and receiver chain, we apply the \ac{mts} of RFSoC to calibrate the random and unknown delay between different RFDC interfaces. 

In this work, the obtained \ac{ra} tensors are continuously streamed to the host PC for further processing. Our first goal is to detect and track possible targets in the environment. This poses different challenges, such as strong self-interference and background interference, as well as multi-target association, which are addressed in our designed algorithms.


\section{Algorithms}
\begin{figure*}[t] 
    \centering
    \includegraphics[width=\textwidth]{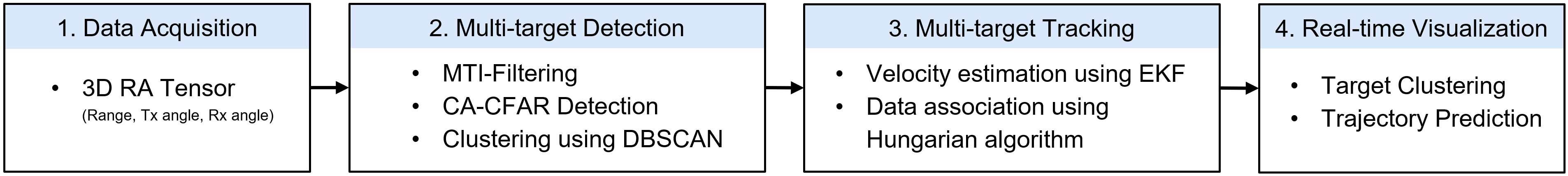}
    \caption{Real-time multi-target detection and tracking pipeline.}
    \label{fig1_pipeline}
\end{figure*}

To efficiently generate target tracks from the RA tensors, we implement a multi-stage pipeline as shown in Fig.~\ref{fig1_pipeline}, which is divided into two main parts: target detection and target tracking. 

\subsection{Target Detection}

For each sensing sweep, a raw 3D RA tensor is obtained, which captures the backscattered signal power across all combinations of transmit and receive angles. Prior to detection, this tensor is processed using a temporal moving target indication (MTI) filter to suppress static clutter and strong self-interference. A high-pass finite impulse response (FIR) filter is applied along the slow-time dimension of the tensor, effectively removing static objects and gradually absorbing very slow-moving targets into the background. This MTI-based adaptive background subtraction significantly improves the signal-to-clutter ratio. 

Using the resulting filtered data that is cleaner and less noisy, we then apply a widely used detection algorithm in radar and sensing systems, the CA-CFAR. This algorithm, also known as adaptive threshold detection \cite{richards2005fundamentals}, estimates the local average noise power and applies a spatially varying threshold to maintain a controlled false alarm rate. We chose CA-CFAR because noise and clutter statistics vary across range and angle, hence determining a single global detection threshold is difficult. This is achieved by generating matrices and tensors containing zeros and ones to indicate the presence or absence of a target, effectively separating it from noise or clutter. Following CA-CFAR, we proceed to perform clustering of the detected points using the DBSCAN algorithm. This algorithm is responsible for clustering spatially neighboring detections into compact groups while naturally rejecting isolated noise points. This step is necessary because multiple CFAR detections within a tensor may originate from a single, extended target.


\subsection{Target Tracking}

In the tracking stage, we apply EKF to estimate the velocity of detected targets over time. The clustered RA measurements obtained from the detection stage are first transformed from the range-angle domain into Cartesian coordinates. The coordinate conversion is given by 
\begin{align}
\begin{bmatrix}
    x \\
    y
\end{bmatrix}
=
\begin{bmatrix}
r\cdot\sin(\theta) \\
r\cdot\cos(\theta)
\end{bmatrix}
\label{eq:equation2}
\end{align}
where $r$ is the range from the mmWave beamformers to the targets and $\theta$ represents the corresponding steering angle obtained from the transmit and receive beam pair. This coordinate definition is chosen such that the cross-range (horizontal) component is represented along the x-axis, while the down-range (depth) component is represented along the y-axis, providing an intuitive Cartesian visualization of the radar scene. 

This coordinate transformation using \eqref{eq:equation2} enables the use of a state vector in Cartesian coordinates. Each target is modeled by a Cartesian state vector comprising position and velocity components. The state vector, $\mathbf{x}_k$, is shown by
\begin{align}
\mathbf{x}_k = \begin{bmatrix} x_k & y_k & \dot{x}_k & \dot{y}_k \end{bmatrix}^\mathrm{T}
\label{eq:equation3}
\end{align}
where $(x_k, y_k)$ denote the target position and $(\dot{x}_k, \dot{y}_k)$ represent its velocity at slow-time index $k$. A motion model with constant velocity is assumed, which is suitable for short-term human and object motion, while unmodeled accelerations are accounted for through process noise.

The EKF operates recursively via two states: prediction and update. In the prediction step, the EKF extrapolates the target state forward in time based on the assumed motion model, providing an estimate of the target's future position and velocity prior to receiving new measurements. In the update step, incoming measurements obtained from the DBSCAN algorithm are used to refine the predicted state estimate. Measurement noise and model uncertainties are also taken into consideration. This algorithm is particularly suited for this application as real-world target movements involve acceleration in every direction, which exhibit non-linear dynamics. The recursive nature of the EKF allows target states to be maintained even during brief periods of missed or unreliable detections by relying on model-based predictions. This improves track continuity and robustness in cluttered environments and under varying measurement quality. Track creation and termination are handled based on the consistency of successive detections over time, helping to filter out false positives while maintaining stable tracks for persistent targets.

In multi-target scenarios, the predicted states obtained from the EKF must be associated with the newly detected measurement clusters to maintain consistent target trajectories over time. From the identified clusters, we can create individual tracks for each target and use a global nearest neighbor data association strategy. This association technique is formulated as a global minimum-cost assignment problem based on the Euclidean distance between predicted EKF track positions and incoming measurement clusters. We apply the Hungarian algorithm to solve this assignment problem, ensuring one-to-one associations and consistent track labeling across frames. The Hungarian algorithm efficiently solves the bipartite assignment problem in polynomial time complexity and guarantees a globally optimal one-to-one association. This global optimization avoids ambiguous or conflicting assignments that may arise from greedy matching strategies, particularly in the presence of closely spaced targets.

\section{Results}
\begin{figure}[t] 
    \centering
    \includegraphics[width=0.48\textwidth]{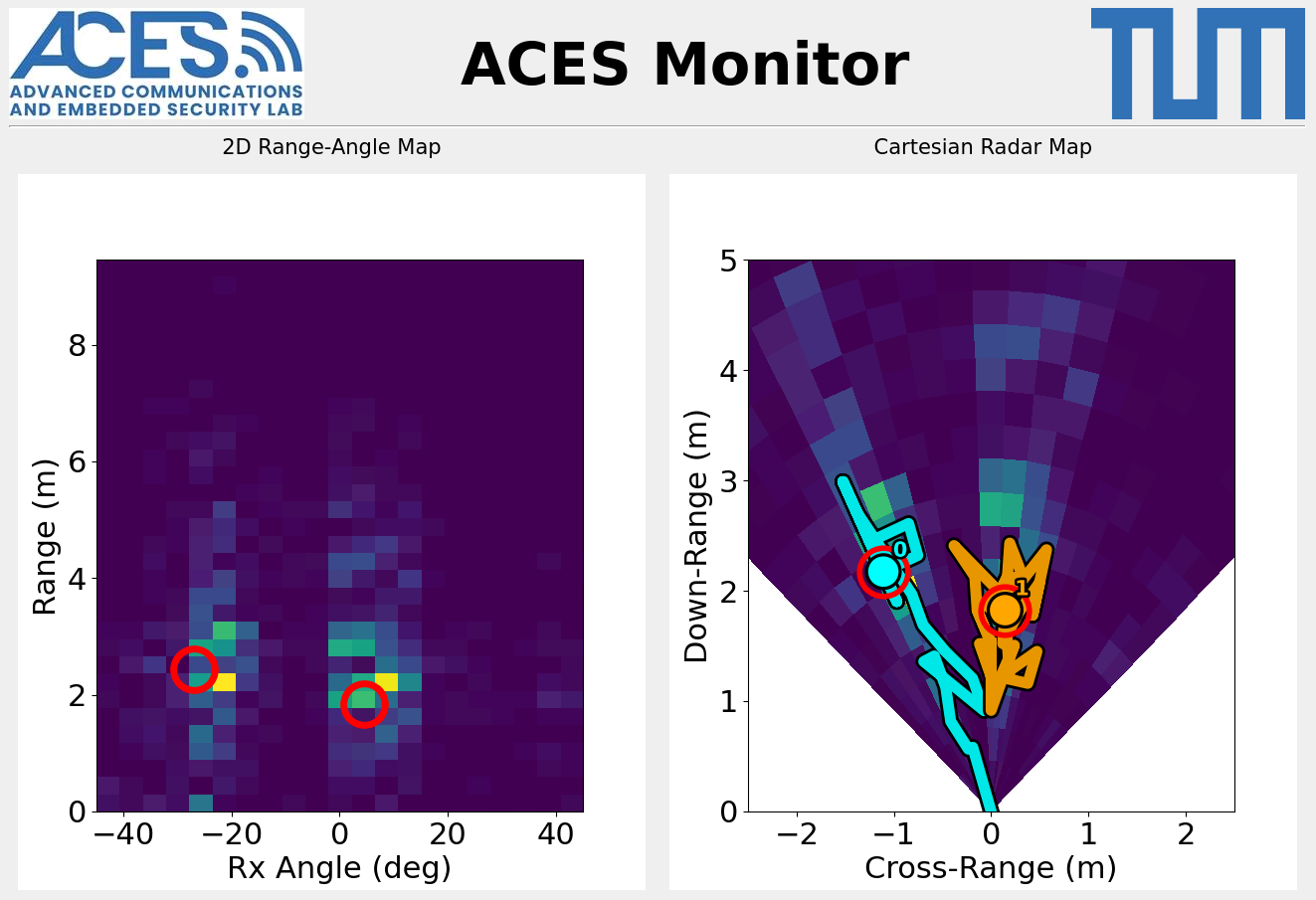}
    \caption{Visualization of detection and tracking results.}
    \label{fig:GUI}
\end{figure}

The system's capabilities are shown using two moving radar reflectors in the environment. Fig.~\ref{fig:GUI} shows a snapshot of the detection and tracking results. The left panel presents a 2D RA map, where target reflections are clearly visible after adaptive clutter subtraction. The right panel illustrates the corresponding Cartesian radar map obtained by transforming the detected measurements into Cartesian coordinates for improved spatial interpretation.

Detected points are clustered using DBSCAN algorithm, with the resulting clusters circled in red. Each cluster is assigned a unique track identifier, labeled as \textit{0} and \textit{1} corresponding to the two distinct reflectors. The associated target tracks are drawn in the Cartesian domain, illustrating the targets' history, demonstrating successful multi-target detection and continuous tracking of two distinct moving reflectors in real-time.



\bibliographystyle{IEEEtran}
\bibliography{IEEEabrv,mybib.bib}
\end{document}